\documentclass[apjl]{emulateapj}
\usepackage{graphicx}
\usepackage{amssymb}
\usepackage{amsmath}
\usepackage{natbib}
\usepackage{color}
\usepackage[flushleft]{threeparttable}

\newcommand\be{\begin{equation}}
\newcommand\ee{\end{equation}}

\begin{document}

\title{Fast and Slow Precession of Gaseous Debris Disks Around Planet-Accreting White Dwarfs}

\author{Ryan Miranda\altaffilmark{1,3} and Roman R. Rafikov\altaffilmark{1,2}}

\altaffiltext{1}{Institute for Advanced Study, Einstein Drive, Princeton, NJ 08540}
\altaffiltext{2}{Centre for Mathematical Sciences, Department of Applied Mathematics and Theoretical Physics, University of Cambridge, Wilberforce Road, Cambridge CB3 0WA, UK}
\altaffiltext{3}{miranda@ias.edu}

\begin{abstract}
Spectroscopic observations of some metal-rich white dwarfs (WDs), believed to be polluted by planetary material, reveal the presence of compact gaseous metallic disks orbiting them. The observed variability of asymmetric, double-peaked emission line profiles in about half of such systems could be interpreted as the signature of precession of an eccentric gaseous debris disk. The variability timescales --- from decades down to $1.4$ yr (recently inferred for the debris disk around HE 1349--2305) --- are in rough agreement with the rate of general relativistic (GR) precession in the test particle limit. However, it has not been demonstrated that this mechanism can drive such a fast, coherent precession of a radially extended (out to $1 R_\odot$) gaseous disk mediated by internal stresses (pressure). Here we use the linear theory of eccentricity evolution in hydrodynamic disks to determine several key properties of eccentric modes in gaseous debris disks around WDs. We find a critical dependence of both the precession period and radial eccentricity distribution of the modes on the inner disk radius, $r_\mathrm{in}$. For small inner radii, $r_\mathrm{in} \lesssim (0.2 - 0.4) R_\odot$, the modes are GR-driven, with periods of $\approx 1 - 10$ yr. For $r_\mathrm{in} \gtrsim (0.2 - 0.4) R_\odot$, the modes are pressure-dominated, with periods of $\approx 3 - 20$ yr. Correspondence between the variability periods and inferred inner radii of the observed disks is in general agreement with this trend. In particular, the short period of HE 1349--2305 is consistent with its small $r_\mathrm{in}$. Circum-WD debris disks may thus serve as natural laboratories for studying the evolution of eccentric gaseous disks.
\end{abstract}

\keywords{accretion, accretion disks --- hydrodynamics --- white dwarfs --- relativistic processes}


\section{Introduction}


Tens of percent of white dwarfs (WDs) show signs of metal pollution \citep{Farihi2016}, which is generally believed to be caused by the accretion of high-Z material originating from circum-WD planetary systems \citep{Debes,Jura2003}. A number of such WDs exhibit infrared (IR) excesses \citep{Jura2003,Jura2007,Farihi2016}, revealing the presence of compact ($\lesssim R_\odot$), warm ($T\sim 500-2000$ K) and dense disks orbiting these stellar remnants. Moreover, eight metal-rich WDs with IR excesses also show double-peaked metal {\it emission} lines, indicative of high metallicity compact {\it gaseous} disks in Keplerian rotation around them  \citep{Gan2006,Gan2007,Gan2008,Gan2011,Melis2012,Farihi0959}. Three of these systems exhibit {\it roughly periodic} time variability\footnote{\citet{Wilson2014} reported variation of the {\it strength} of the lines in SDSS J1617+1620, culminating in their disappearance.} of the emission line profiles of the Ca \textsc{ii} triplet, with periods of one to several decades \citep{Wilson,Manser1043,Manser1228}. Additionally, the gas disk around WD 1145+017 \citep{Xu}, which is also orbited by transiting, disintegrating planetesimals \citep{Vander}, shows periodic variability of {\it absorption} lines (due to Ni \textsc{ii}, Mg \textsc{i}, and Fe \textsc{ii}) with a period of $5.3$ years \citep{Red,Cauley}; properties of circum-WD disks showing quasi-periodic variability are summarized in Table \ref{table:properties}. 

\citet{Manser1228} suggested that emission line variability is the signature of an {\it eccentric, precessing} gas disk, and that the variability periods are broadly consistent with the general relativistic (GR) precession of a {\it test particle} with a semi-major axis comparable to the stellocentric radii from which gas emission is detected. However, real gaseous disks are {\it fluid} entities, meaning that understanding their precession requires a full hydrodynamic treatment.

Recently, \citet{d18} reported rapid variability of the Ca \textsc{ii} triplet emission from HE 1349--2305 with a period of $1.4$ years, an order of magnitude shorter than in other WD debris disks. This variability has been reported to be inconsistent with GR precession, due to the large disparity (factor of $\approx 50$) between the GR precession rate at the inner and outer disk edges ($0.2$ and $1 R_\odot$).

In this paper, we use the linear theory for the evolution of eccentric disks \citep{o01,to16} to model the hydrodynamic behavior of eccentric gaseous debris disks orbiting WDs. In general, an eccentric disk can be described by a series of global modes, each with a corresponding radial eccentricity profile and coherent precession frequency, which we compute in this work. We demonstrate that the location of the inner edge of the disk plays a critical role in setting the global disk precession period and can explain the range of observed variability periods. We also show that the rapid variability of HE 1349--2305 is consistent with its small inner radius.


\section{Eccentric Disk Dynamics}


Thin fluid disks can in general be eccentric, and may precess due to external forces or internal stresses \citep[e.g.][]{o01,Statler}. Our goal is to understand the precession periods of eccentric WD debris disks, to determine whether or not they can be identified with the observed periods of variability. We therefore do not attempt to address the process by which these disks become eccentric, or how the eccentricity is maintained over long (relative to the precession period) timescales (if it is in fact maintained over such timescales). Rather, we simply assume that the disk has somehow acquired a significant eccentricity (see section \ref{sec:discussion} for a more detailed discussion), and seek to understand its subsequent behavior.

The structure of an eccentric disk is described by the eccentricity $e(r,t)$ and argument of pericenter $\varpi(r,t)$, which are combined in the complex disk eccentricity
\be
E(r,t) = e\exp(\mathrm{i}\varpi).
\ee

\citet{o01} developed a disk eccentricity evolution theory, which has been applied to a variety of astrophysical systems, including accreting black holes \citep{fo09}, Be stars \citep{Ogilvie2008}, and protoplanetary disks with embedded giant planets \citep{to16}. In the framework of this theory, the linear ($e \ll 1$) equation describing the evolution of $E(r,t)$ for a non-self-gravitating, locally isothermal disk is \citep{to16}
\be
\label{eq:E_evolution}
\begin{aligned}
-\frac{2\mathrm{i}\Omega}{c_\mathrm{s}^2}\frac{\partial E}{\partial t} & = \frac{\partial^2 E}{\partial r^2} + \left(\frac{3}{r} + \frac{\mathrm{d}\ln\Sigma}{\mathrm{d}r}\right)\frac{\partial E}{\partial r} \\
& + \left[\frac{1}{r}\frac{\mathrm{d}\ln\Sigma}{\mathrm{d}r} + \frac{\mathrm{d}\ln c_\mathrm{s}^2}{\mathrm{d}r}\left(\frac{1}{r} - \frac{\mathrm{d}\ln\Sigma}{\mathrm{d}r}\right) \right. \\
& \left. - \frac{1}{c_\mathrm{s}^2}\frac{\mathrm{d}^2 c_\mathrm{s}^2}{\mathrm{d}r^2} + \frac{6}{r^2} + \frac{2\Omega}{c_\mathrm{s}^2}\dot{\varpi}_\mathrm{GR}\right]E,
\end{aligned}
\ee
where $\Sigma(r)$ is the disk surface density, $c_\mathrm{s}(r)$ is the sound speed, and $\Omega(r) = (GM_*/r^3)^{1/2}$ is the Keplerian orbital frequency. Equation (\ref{eq:E_evolution}) includes the terms (most importantly $6r^{-2}E$) describing the 3D effect related to the variation of the vertical gravitational force exerted on a fluid element as it moves along an eccentric orbit \citep{Ogilvie2008}, which has a significant impact on our results (see section \ref{sec:discussion}). We have also added the last term multiplying $E$ on the right-hand side of equation (\ref{eq:E_evolution}) to describe GR precession with the frequency (in the limit $e \ll 1$)
\be
\begin{aligned}
\dot{\varpi}_\mathrm{GR} & = \frac{3GM_*}{c^2 r} \Omega \\
& = \frac{2\pi}{107 ~\mathrm{yr}} \left(\frac{M_*}{0.6 M_\odot}\right)^{3/2} \left(\frac{r}{R\odot}\right)^{-5/2},
\end{aligned}
\label{eq:GR}
\ee
where $c$ is the speed of light.

In equation (\ref{eq:E_evolution}) we neglect terms associated with viscosity, $\nu = \alpha c_\mathrm{s}^2/\Omega$ \citep{o01}, where $\alpha$ is the dimensionless effective viscosity \citep{SS}. This is justified since (1) we do not consider excitation/damping of the disk eccentricity and (2) the characteristic viscous time
\be
t_\nu\sim \frac{r^2}{\nu}\approx 600~\mbox{yr}\left(\frac{r}{R_\odot}\right)^{1/2}\left(\frac{10^{-2}}{\alpha}\right)\left(\frac{5000~\mbox{K}}{T}\right)
\label{eq:t_nu}
\ee
(for $M_\star=0.6M_\odot$ and the mean molecular weight of the gas $\mu=28m_p$) is significantly longer than the observed variability periods (or theoretical eccentricity evolution timescale). 

The terms inside the brackets multiplying $E$ in equation (\ref{eq:E_evolution}) represent different sources of differential precession. All of them except the last one characterize effects related to the disk pressure, which together typically result in prograde precession.\footnote{Pressure leads to prograde precession because of the significant role of the term related to the 3D effect described by \citet{Ogilvie2008}. If this effect is neglected, then pressure typically leads to retrograde precession \citep{go06}.} The last term describes GR-driven precession, which is always prograde. As the two types of terms scale differently with $r$, there is a critical radius, $r_\mathrm{crit}$, at which they become comparable in magnitude, which delineates the region in which GR is dominant ($r < r_\mathrm{crit}$), from the region in which pressure is dominant ($r > r_\mathrm{crit}$). Assuming the disk surface density and sound speed are described by
\be
\Sigma = \Sigma_\mathrm{in} \left(\frac{r}{r_\mathrm{in}}\right)^{-p}
\label{eq:Sigma}
\ee
and
\be
c_\mathrm{s}^2 = \frac{kT_\mathrm{in}}{\mu} \left(\frac{r}{r_\mathrm{in}}\right)^{-q},
\label{eq:T}
\ee
where $T_\mathrm{in}$ and $\Sigma_\mathrm{in}$ are the gas temperature and surface density\footnote{Note that $\Sigma_\mathrm{in}$ drops out of equation (\ref{eq:E_evolution}).} at $r_\mathrm{in}$ ($p$ and $q$ are the constant power law indices), the critical radius is given by
\be
\label{eq:rcrit}
r_\mathrm{crit} = \left[\frac{\beta c^2 k T_\mathrm{in} r_\mathrm{in}^q}{6\mu (GM_*)^2}\right]^{1/(q-2)},
\ee
where $\beta = |6-q(q+2)-p(q+1)|$. For a globally isothermal disk ($q = 0$),
\be
\begin{aligned}
r_\mathrm{crit} & = 0.38 R_\odot \left(\frac{6-p}{4}\right)^{-1/2}  \left(\frac{T}{5000 \mathrm{K}}\right)^{-1/2} \\
& \times \left(\frac{\mu}{28 m_\mathrm{p}}\right)^{1/2} \left(\frac{M_*}{0.6 M_\odot}\right).
\label{eq:rcrit_est}
\end{aligned}
\ee
The gaseous component of a typical WD debris disk can conceivably lie entirely within the pressure-dominated region, entirely within the GR-dominated region, or span both regions, depending on its radial extent. The dominant physical mechanism responsible for the precession of such a disk is then determined by the details of its structure. 


\subsection{Normal Modes}
\label{subsec:modes}


We look for the normal mode solutions of equation (\ref{eq:E_evolution}) of the form
\be
E(r,t) = E(r)\exp(\mathrm{i}\omega_\mathrm{prec} t),
\ee
so that the entire disk precesses coherently with an angular frequency $\omega_\mathrm{prec}$. Equation (\ref{eq:E_evolution}) then becomes an ordinary differential equation for $E(r)$:
\be
\label{eq:E_mode}
\begin{aligned}
& \frac{\partial^2 E}{\partial r^2} + \frac{(3-p)}{r}\frac{\partial E}{\partial r} \\
+ & \left[\frac{6-q(q+2)-p(q+1)}{r^2} + \frac{6r^2\Omega^4}{c^2 c_\mathrm{s}^2} - \frac{2\Omega\omega_\mathrm{prec}}{c_\mathrm{s}^2}\right]E = 0.
\end{aligned}
\ee
We solve the eigenvalue equation (\ref{eq:E_mode}), supplied with a choice of boundary conditions (BCs), using the shooting method. The solutions constitute a spectrum of eigenvalues $\omega_\mathrm{prec}$ and associated eigenfunctions $E(r)$, each with a different number of nodes --- radial locations at which $|E| = 0$ and $\varpi$ experiences a $180^\circ$ shift. Note that $E(r)$, which refers to linear mode solution, is distinct from the physical disk eccentricity $e(r)$, which differs by an amplitude factor (see section \ref{subsec:mode_amplitude}).

For the inner and outer BCs, we choose
\be
\frac{\partial E(r_\mathrm{in})}{\partial r} = \frac{\partial E(r_\mathrm{out})}{\partial r} = 0.
\label{eq:BC}
\ee
The choice of the inner BC is motivated by the observed emission line profiles. A disk with a circular inner edge should produce double-peaked lines, equidistant from the rest frame velocity and with equal height. The observed line profiles show a distinct difference between the maximum redshifted and blueshifted velocities of the line peaks (as well as different heights), indicating that the inner edge of the disk has a significant eccentricity. The zero-gradient BC applied to $E$ at $r_\mathrm{in}$ allows the inner disk edge to be eccentric (as opposed to, e.g., setting $E = 0$ at $r_\mathrm{in}$). Also, we found the outer BC to be relatively unimportant in determining the mode properties, so we simply apply the same BC at $r_\mathrm{out}$.

For several reasons, we focus only on the lowest-order mode, whose eigenfunction has no nodes. First, modeling of the disk eccentricity distribution for WD 1145+017 by \citet{Cauley} indicates that $e$ varies on a length scale comparable to the disk radius, which is indicative of a low-order mode. Second, for higher-order modes, the disk gets divided into multiple (but smoothly connected) eccentric, anti-aligned sub-disks. As a result, the systematic difference in orbital velocities from one side of the disk to the other (for some range of $r$) is reduced, suppressing the asymmetry of the double-peaked emission lines. Third, the lowest-order mode typically has the longest precession period, in a way setting an upper limit on the precession period. Finally, the lowest-order mode is typically the least affected by viscous damping, when this effect is considered \citep{go06}.

We consider a range of values for $r_\mathrm{in}$ and $r_\mathrm{out}$ (motivated by the actual measurements in disk-hosting systems, see Table \ref{table:properties}), inner disk temperature $T_\mathrm{in}$, and for the surface density and temperature power-law indices $p$ and $q$. Under the assumption that the gas disk is fed by the sublimation of a particulate debris disk \citep{Rafikov1,Rafikov2} at the sublimation radius $\sim 0.2R_\odot$ \citep{Garmilla}, one can show that the $\Sigma$ profile with $p = 2$ should naturally develop in a globally isothermal disk outside this radius \citep{Metzger,Rafikov16}. However, we also look at profiles with $p = 1$, which have more mass at large radii. The thermal structure of gaseous circum-WD disks was computed in \citet{Melis2010}, who showed that around hot WDs, $T_\mathrm{in}\sim 10^4$ K with $q \approx 0.5 - 1$ may be typical. However, for completeness we also consider the possibility of colder disks and lower $q = 0$.


\section{Results}


\begin{figure}
\begin{center}
\includegraphics[width=0.49\textwidth,clip]{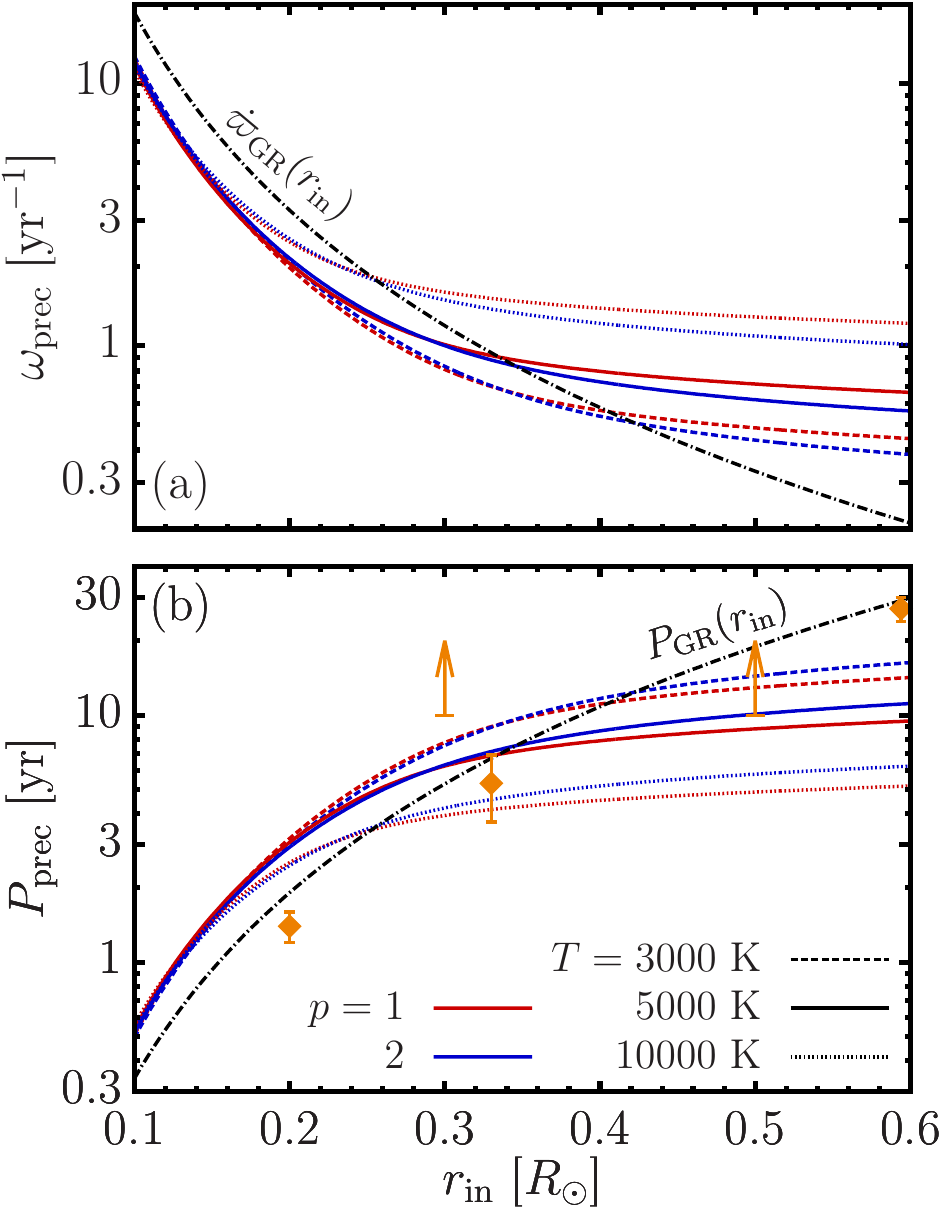}
\caption{Mode precession frequency, $\omega_\mathrm{prec}$ (top panel; a), and precession period, $P_\mathrm{prec} = 2\pi/|\omega_\mathrm{prec}|$ (bottom panel; b), as a function of the inner disk radius $r_\mathrm{in}$, for disk models with different values for the (radially constant, $q=0$) gas temperature $T$ and surface density power law index $p$. The outer disk radius $r_\mathrm{out}$ is fixed at $1 R_\odot$ and the WD mass is $0.6 M_\odot$. For reference, the dot-dashed curves indicate the GR precession frequency and period for a nearly circular test particle at $r_\mathrm{in}$. The orange symbols indicate the inferred $r_\mathrm{in}$ and $P_\mathrm{prec}$ of the observed variable debris disks. Uncertainties in the measured precession periods are shown where available (a $30\%$ uncertainty has been adopted in the case of WD 1145+017), and lower limits are shown for cases in which a definite period has not yet been determined (see Table \ref{table:properties}).}
\label{fig:freq_period}
\end{center}
\end{figure}

\begin{figure}
\begin{center}
\includegraphics[width=0.49\textwidth,clip]{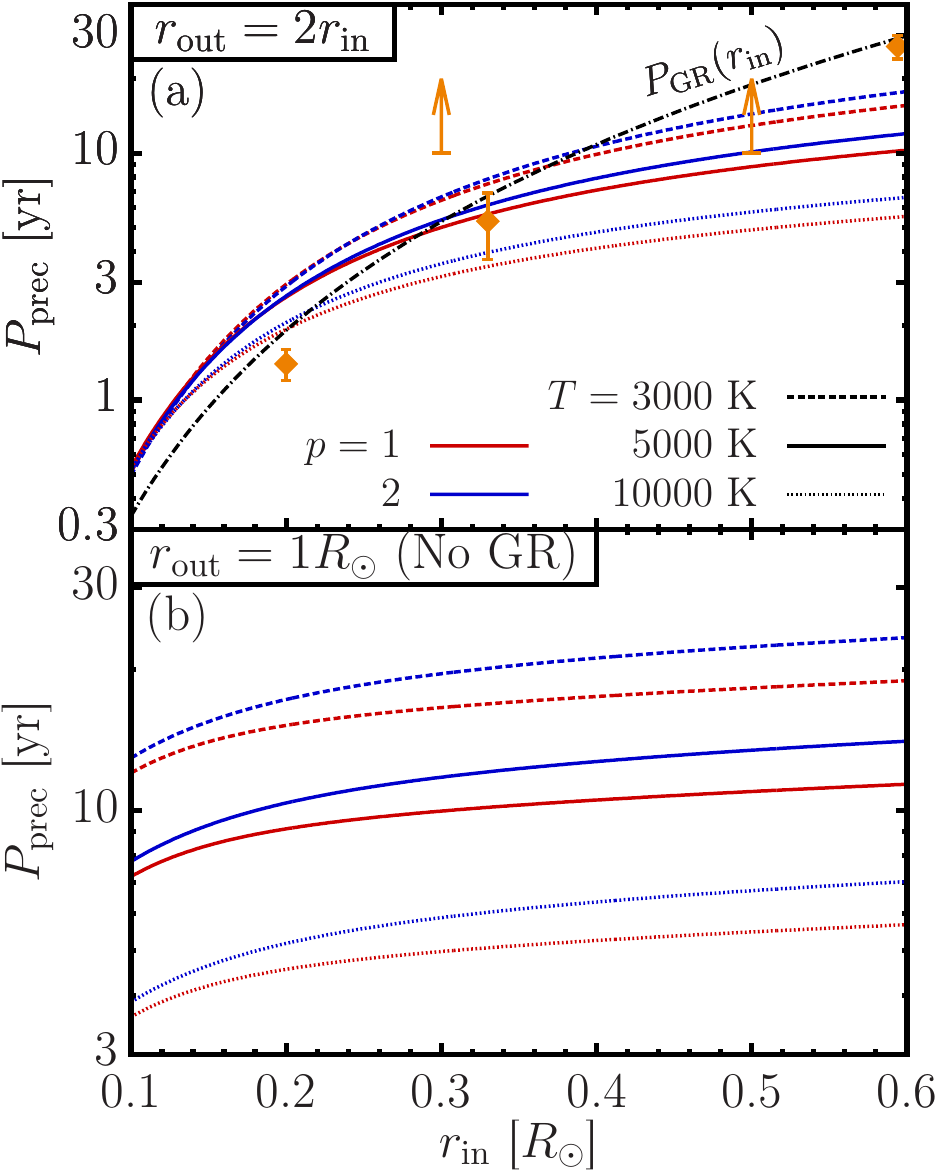}
\caption{Mode precession period (as in Figure \ref{fig:period_cases}b) for the case of a narrow disk with $r_\mathrm{out} = 2 r_\mathrm{in}$ (top panel; a), and for a disk with $r_\mathrm{out} = 1 R_\odot$, but ignoring the effect of GR, so that the modes are purely pressure-dominated (bottom panel; b). In (a), the inferred $r_\mathrm{in}$ and $P_\mathrm{prec}$ of the observed variable debris disks are also shown.}
\label{fig:period_cases}
\end{center}
\end{figure}

\begin{figure}
\begin{center}
\includegraphics[width=0.49\textwidth,clip]{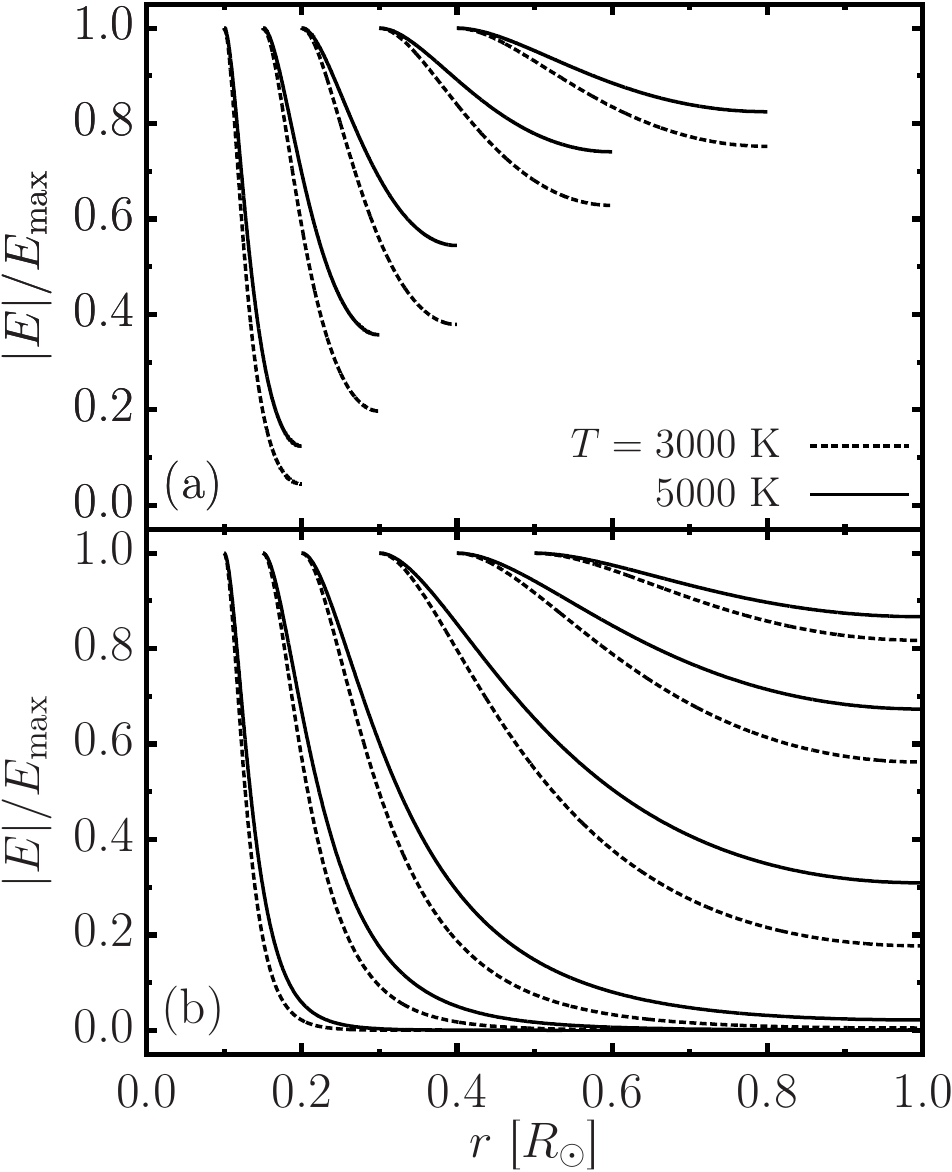}
\caption{Mode eccentricity profile (normalized to unity) for several different inner disk radii and temperatures, for disks with $r_\mathrm{out} = 2 r_\mathrm{in}$ (top panel; a) and with $r_\mathrm{out} = 1 R_\odot$ (bottom panel; b). All models shown have a surface density power law index $p = 2$.}
\label{fig:modes}
\end{center}
\end{figure}

\begin{figure}
\begin{center}
\includegraphics[width=0.49\textwidth,clip]{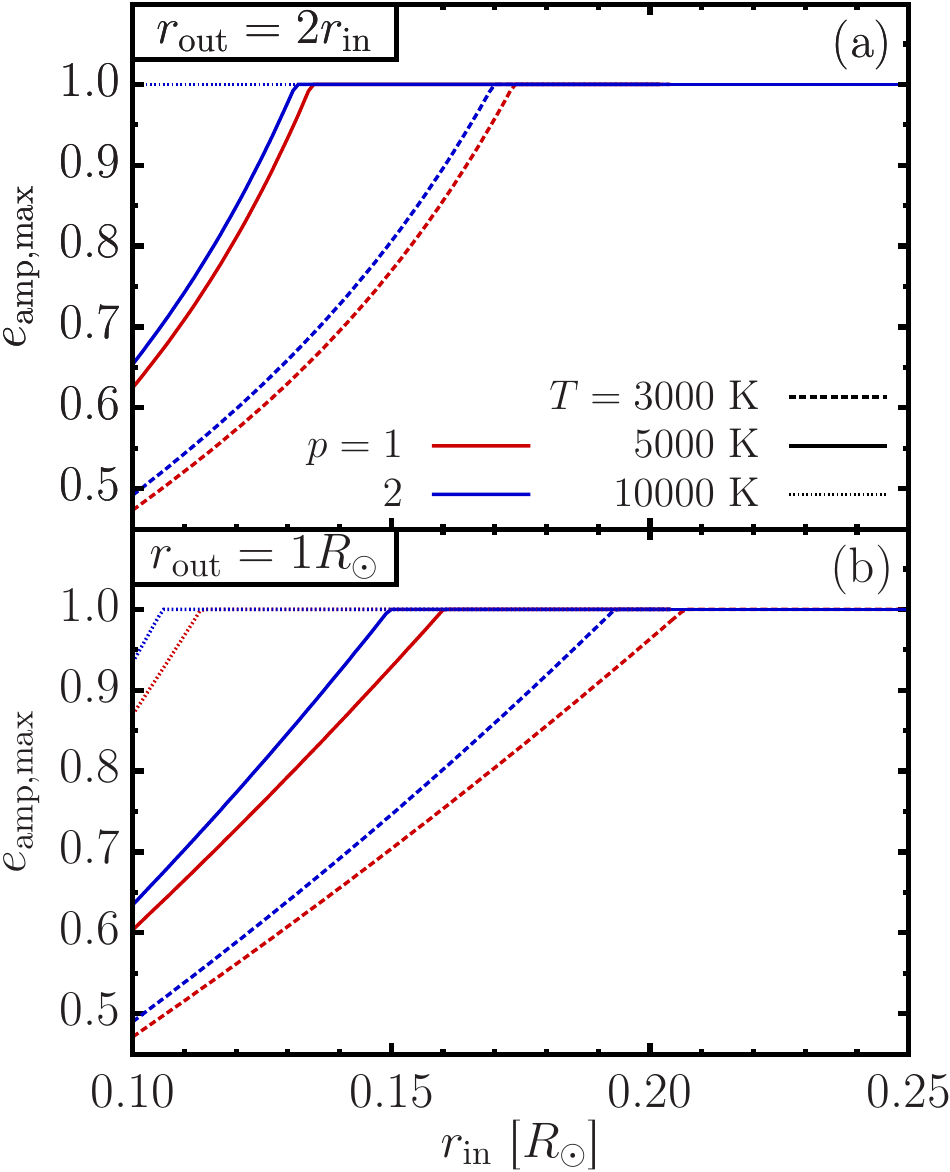}
\caption{Maximum mode amplitude, $e_\mathrm{amp,max}$, (see equation \ref{eq:max_amp}) as a function of $r_\mathrm{in}$ for a variety of disk models, for the case of a narrow disk with $r_\mathrm{out} = 2 r_\mathrm{in}$ (top panel; a) and an extended disk with $r_\mathrm{out} = 1 R_\odot$ (bottom panel; b).}
\label{fig:emax}
\end{center}
\end{figure}

The mode precession frequency, $\omega_\mathrm{prec}$, and precession period, $P_\mathrm{prec} = 2\pi/|\omega_\mathrm{prec}|$, are shown in Figure \ref{fig:freq_period} as a function of inner disk radius $r_\mathrm{in}$ for a variety of disk models with $r_\mathrm{out}$ fixed at $1 R_\odot$. We focus on the case of a globally isothermal ($q = 0$) disk, since we found the mode properties to depend only weakly on the slope of the temperature profile (at least for $0 < q < 1/2$). Note that $\omega_\mathrm{prec}$ is always positive, corresponding to prograde precession, and that $P_\mathrm{prec}$ increases with $r_\mathrm{in}$, but much more steeply for $r_\mathrm{in} \lesssim r_\mathrm{crit}$ than for $r_\mathrm{in} \gtrsim r_\mathrm{crit}$, where it becomes almost constant.

For $r_\mathrm{in} \lesssim r_\mathrm{crit}$, the modes are GR dominated, with periods of $\approx 1 - 10$ years. In this regime, $P_\mathrm{prec}$ is about twice as large as  $P_\mathrm{GR}(r_\mathrm{in})$, the GR precession period of a nearly-circular test particle at $r_\mathrm{in}$. For very small $r_\mathrm{in}$ ($\lesssim 0.15 R_\odot$), a lower disk temperature results in a smaller $P_\mathrm{prec}$ [closer to $P_\mathrm{GR}(r_\mathrm{in})$] for a given $r_\mathrm{in}$, since hotter disks transition to the pressure-dominated regime at smaller $r_{\rm in}$ (see equation \ref{eq:rcrit_est}). Colder disks are therefore forced to precess at a rate closer to the fast $\dot\varpi_\mathrm{GR}(r_\mathrm{in})$, suggesting that $P_\mathrm{prec} \to P_\mathrm{GR}(r_\mathrm{in})$ in the limit $c_\mathrm{s} \to 0$. 

For $r_\mathrm{in} \gtrsim r_\mathrm{crit}$, the modes are pressure-dominated and have periods of $\approx 3 - 20$ years [notice that $P_\mathrm{prec} < P_\mathrm{GR}(r_\mathrm{in})$ in this regime]. Periods in this range arise for purely pressure-dominated disks, when the effect of GR is ignored. This is illustrated in Figure \ref{fig:period_cases}b, where $P_\mathrm{prec}$ is calculated by dropping the GR precession term in equation (\ref{eq:E_mode}). Note a very weak dependence of the precession period on $r_\mathrm{in}$ in the case of pure pressure modes. 

The correspondence of $P_\mathrm{prec}$ with $P_\mathrm{GR}(r_\mathrm{in})$ for small $r_\mathrm{in}$ and with the pressure-driven precession period for large $r_\mathrm{in}$ is not sensitive to the outer disk radius. Figure \ref{fig:period_cases}a demonstrates this by showing that the $P_\mathrm{prec}(r_\mathrm{in})$ profile for narrow disks with $r_\mathrm{out} = 2 r_\mathrm{in}$ is not very different from Figure \ref{fig:freq_period}b. Nor is the qualitative $P_\mathrm{prec}(r_\mathrm{in})$ behavior sensitive to the disk surface density profile (see Figures \ref{fig:freq_period}b, \ref{fig:period_cases}a), in particular, whether the disk mass is concentrated at the outer edge (for $p = 1$) or is evenly spread in $\log r$ (as for $p = 2$).

In Figure \ref{fig:modes} we show the radial eccentricity profiles of the modes $E(r)$ computed for disks with different inner radii $r_\mathrm{in}$ (and different temperatures). For the BCs (\ref{eq:BC}), the maximum eccentricity always occurs at $r_\mathrm{in}$, and $E(r)$ decreases with $r$. For sufficiently small $r_\mathrm{in}$ ($\lesssim 0.3 R_\odot$), in the GR-dominated regime, $E(r)$ sharply decreases near the inner edge of the disk, by at least an e-fold between $r_\mathrm{in}$ and $2 r_\mathrm{in}$. The eccentricity varies much more slowly with $r$ when precession is dominated by pressure, for $r_\mathrm{in}\gtrsim 0.4 R_\odot$. Also, the steepness of the $E(r)$ profile decreases for hotter disks, and is only weakly dependent on the slope of the surface density profile.


\subsection{Mode Amplitude}
\label{subsec:mode_amplitude}


Equations (\ref{eq:E_evolution}) and (\ref{eq:E_mode}) are linear in $E$ and, thus, cannot predict the {\it amplitude} of the mode; they yield only the radial profile of $E$. In real disks the mode amplitude is ultimately determined by the balance of eccentricity excitation and damping, which we do not address. Nevertheless, we can still come up with an {\it upper limit} on the mode amplitude, which in some cases can be more restrictive than the obvious condition $e < 1$.

Indeed, to be physically realizable, the eccentricity profile must satisfy \footnote{Formally, the eccentricity profile is a function of semi-major axis $a$. In the linear framework for which the modes are computed, $a$ and $r$ are interchangeable. For non-small eccentricity, $e(r)$ should be thought of as being representative of $e(a)$.}
\be
\label{eq:orbitcross}
\left|e(a)+\frac{\mathrm{d}e(a)}{\mathrm{d}\ln a}\right| < 1,
\ee
otherwise adjacent orbits would cross one another \citep{o01,Statler}. This requires $e$ to be less than the limiting eccentricity,
\be
e<e_\mathrm{lim}(r) = \left|1+\frac{\mathrm{d}\ln e(r)}{\mathrm{d}\ln r}\right|^{-1},
\ee
for all $r$. 

The physical eccentricity profile of the disk is given by the linear mode eigenfunction $E(r)$ obtained from equation (\ref{eq:E_mode}) scaled by an amplitude $e_\mathrm{amp}$: 
\be
e(r) = e_\mathrm{amp} \frac{|E(r)|}{E_\mathrm{max}},
\ee
where $E_\mathrm{max}$ is the maximum value of $|E(r)|$, which in our case occurs at $r = r_\mathrm{in}$, see Figure \ref{fig:modes} [thus $e(r_\mathrm{in}) = e_\mathrm{amp}$]. The maximum mode amplitude for which orbit crossing is guaranteed to be avoided is then 
\be
\label{eq:max_amp}
e_\mathrm{amp,max} = \mathrm{min}\left[e_\mathrm{lim}(r)\frac{E_\mathrm{max}}{|E(r)|}\right],
\ee
where we minimize over the full radial extent of the disk and use $E(r)$ and $E_\mathrm{max}$ from our linear mode calculation.

The results of such a calculation are shown in Figure \ref{fig:emax}. The limitation (\ref{eq:orbitcross}) turns out to be only weakly restrictive when the BCs (\ref{eq:BC}) are used; the mode can in principle take on any amplitude (less than unity) unless $r_\mathrm{in} \lesssim (0.15 - 0.2) R_\odot$, in which case equation (\ref{eq:max_amp}) is modestly restrictive, requiring $e_\mathrm{amp} \lesssim 0.6$ for $r_\mathrm{in} \approx 0.1 R_\odot$. This amplitude is large compared to observationally inferred values of $e \approx 0.02$ for SDSS 1228 \citep{Gan2006}, and $e \approx 0.25 - 0.30$ for WD 1145 \citep{Cauley}. If instead we were to choose $E(r_\mathrm{in}) = 0$ as our inner BC, then the maximum mode amplitudes would be about three times smaller (note however that such a BC is disfavored by observations; see section \ref{subsec:modes}).


\section{Discussion}
\label{sec:discussion}


\begin{table*}
\caption{Properties of WDs with time-varying gaseous debris disks}
\begin{tabular}{l|ccccc}
    \hline \hline\\
    Object & HE 1349--2305 & SDSS J1228+1040 & SDSS J0845+2257 (Ton 345) & SDSS J1043+0855 & WD 1145+017 \\ \\
    \hline \hline 
    Type & DA & DA & DB & DA & DB \\ 
    $T_{\rm eff}, K$ & 18,000 & 20,700 & 19,800 & 17,900 & 15,900 \\ 
    $M_\star$, $M_\odot$ & 0.67 & 0.7 & 0.68 & 0.69 & 0.6 \\ 
    $R_\star$, $R_\odot$ & 0.011 & 0.011 & 0.011 & 0.011 &  0.013 \\ 
    $\dot M_Z$, $10^8$ g s$^{-1}$ & 1.3 & 5.6 & 160 &  2.5-12 &  430 \\ 
    \hline
    Lines showing variability\footnotemark[1]{} & e & e & e & e & a  \\ 
    Gas disk & & & & &   \\ 
    ~~~~~~$r_{\rm in}$, $R_\odot$ & 0.2 & 0.6 & 0.5 & $\sim 0.3$ & 0.33  \\ 
    ~~~~~~$r_{\rm out}$, $R_{\odot}$ & 1 & 1.2 & 1 & 0.9  & 0.52  \\ 
    ~~~~~~$B_{\rm crit}$\footnotemark[2]{}, G & 50 & 750 & 2,880 & 150-320  & 1,340  \\ 
    Dust disk\footnotemark[3]{} & & & & &   \\ 
    ~~~~~~$r_{\rm in}$, $R_\odot$ & 0.15 & 0.28 & 0.17 & 0.23 & $\sim 0.26$ \\ 
    ~~~~~~$r_{\rm out}$, $R_{\odot}$ & 0.7 & 1 & 0.9 & 0.49 & (?)  \\ 
    \hline
    Observed precession period, yr & $1.4 \pm 0.2$ & $27 \pm 3$ & $\gtrsim 10$ & $\gtrsim 10$ & 5.3 \\ 
    
    Theoretical precession period\footnotemark[4]{}, yr & 2.5 & 12 & 10 & 5.6 & 5.4 \\
    
    $P_{\rm GR}(r_{\rm in})$\footnotemark[5]{}, yr & 1.6 & 24 & 16 & 4.3 & 5.3 \\ 
    \hline
    References\footnotemark[6]{} & 5,7,14 & 1,4,12 & 3,4,6,8  & 2,4,6,11 & 9,10,13  \\ 
    \hline   
\end{tabular}
\footnotetext[1]{Type of spectroscopic lines used to infer variability of the disk in a given object: e - emission lines, a - absorption lines.} 
\footnotetext[2]{Strength of the WD surface magnetic field necessary to disrupt the gaseous disk accreting at the rate $\dot M_Z$ at a radius $r_{\rm in}$ \citep{Konigl}.} 
\footnotetext[3]{Dust disk radii are very uncertain; inner radii are highly degenerate with the disk inclination \citep{Bergfors}.} 
\footnotetext[4]{Precession period of the linear eccentric mode computed in this work for each system. The mode is computed using the spectroscopically inferred $r_{\rm in}$ and $r_{\rm out}$ of the gas, and assuming a globally isothermal ($q = 0$) disk with a temperature $T = 5000$~K and surface density power law index $p = 2$. See Figures \ref{fig:freq_period}b and \ref{fig:period_cases}a for the effect of varying the disk temperature and surface density profile. Note that this precession period results from the combined effects of pressure and GR, see Figure \ref{fig:period_cases}b for the periods of hypothetical pressure-only modes.}
\footnotetext[5]{Period of the GR precession of a test particle on a nearly-circular orbit, evaluated at the inner radius of the gaseous disk $r_{\rm in}$ inferred from spectroscopic observations.} 
\footnotetext[6]{Key to references: $^{1}$\citet{Gan2006}, $^{2}$\citet{Gan2007}, $^{3}$\citet{Gan2008}, $^{4}$\citet{Melis2010}, $^{5}$\citet{Melis2012}, $^{6}$\citet{Brink}, $^{7}$\citet{Girven}, $^{8}$\citet{Wilson}, $^{9}$\citet{Vander}, $^{10}$\citet{Xu}, $^{11}$\citet{Manser1043}, $^{12}$\citet{Manser1228}, $^{13}$\citet{Cauley}, $^{14}$\citet{d18}.} 
\label{table:properties}
\end{table*}

The most important result of our calculations is finding that the inner disk radius, $r_\mathrm{in}$, plays a decisive role in setting the global precession period of the disk, $P_\mathrm{prec}$. If the inner edge of the disk is inside the GR-dominated region (see equation \ref{eq:rcrit}), with $r_\mathrm{in} \lesssim (0.2 - 0.4) R_\odot$, then we find $P_\mathrm{prec} \approx 1 - 10$ years, about twice as large as $P_\mathrm{GR}(r_\mathrm{in})$, the GR precession period of a test particle at $r_\mathrm{in}$, with shorter periods corresponding to smaller inner radii. For larger $r_\mathrm{in}$, the precession is dominated by pressure, with a period primarily determined by the temperature and surface density profile of the disk. The resulting $P_\mathrm{prec} \approx 3 - 20$ years is only weakly dependent on $r_\mathrm{in}$. Note that periods less than a few years are only possible for $r_\mathrm{in} \lesssim 0.3 R_\odot$, and result from GR-dominated modes.

Looking at the properties of the systems listed in Table \ref{table:properties} (also see Figures \ref{fig:freq_period}b and \ref{fig:period_cases}a), we see that HE 1349--2305, which has the shortest variability period ($1.4$ years), indeed has the smallest $r_\mathrm{in}\approx 0.2 R_\odot$ inferred from the shape of its emission line profiles. Our calculations give a $P_\mathrm{prec}$ about twice as large for the observed value of $r_\mathrm{in}$, and require a slightly smaller $r_\mathrm{in}$ ($\approx 0.15 R_\odot$) to reproduce the observed period. However, we caution that due to the simplified model (isothermal disk with sharply truncated power-law surface density profile) used in our calculations, discrepancies at this level should not be considered too seriously.

The variability periods of SDSS J1043+0855 and WD 1145+017 are consistent with their slightly larger values of $r_\mathrm{in}$ ($\approx 0.3 R_\odot$), which point towards precession periods of $P_\mathrm{prec} \approx 3 - 10$ years, resulting from modes roughly equally affected by pressure and GR. The large values of $r_\mathrm{in}$ inferred for SDSS J1228+1040 and J0845+2257 ($0.5 - 0.6 R_\odot$) are also consistent with their longer periods ($\gtrsim 20$ yr), resulting from pressure-dominated modes. See Table \ref{table:properties} for a more detailed comparison of the observed precession periods and our computed mode precession periods for each system.
 
It is important to emphasize that our calculations do require the disk to have a relatively {\it sharp} inner edge at $r_\mathrm{in}$: if the disk were to extend smoothly all the way to the WD, our calculations would predict much faster GR-dominated precession than found observationally. The existence of an inner edge requires a physical mechanism responsible for truncating the disk at $r_\mathrm{in}$. If $r_\mathrm{in}$ is set by magnetospheric truncation, then the precession rate should be closely linked to the accretion rate $\dot M_Z$ and the WD magnetic field. Table \ref{table:properties} provides estimates of the surface field $B_{\rm crit}$ necessary for disrupting the gaseous debris disk (and creating a magnetospheric cavity) at the inner radius $r_{\rm in }$ \citep{Konigl}, provided that the WD accretes at the rate $\dot M_Z$. In the case of SDSS J1043-0855, the resultant field strength ($\approx 1.3$ kG) is below the current upper limit $B_\star<3$ kG established in \citet{FarihiB}.

The determination of $r_\mathrm{in}$ is not trivial. For an eccentric disk, the model inferring the innermost semi-major axis must involve some information about the disk eccentricity at the inner edge, which is not easy to obtain observationally.\footnote{Such detailed fitting was attempted in \citet{Cauley}, and we adopt their estimates of the minimum disk periastron and eccentricity to infer the innermost semi-major axis, which we associate with $r_{\rm in}$.} For example, the early, low quality spectral data for SDSS J1043+0855 indicated an extremely small size of the inner cavity of the gaseous disk, $r_{\rm in}\approx 12R_\star$ \citep{Melis2010}. However, subsequent higher quality data \citep{Manser1043} suggests that the emission line splitting is in fact smaller than adopted in \citet{Melis2010}, and the disk is more edge-on, all resulting in larger $r_{\rm in}$, which we estimate to be about $0.3R_\odot$.
 
Furthermore, the existing determinations of $r_\mathrm{in}$ are based on the premise that the truncation radius of the disk surface density profile corresponds to the innermost location where the observed emission lines are produced. This is not necessarily true, as the radial span of the emission region can be determined, e.g., by the excitation conditions of the line-emitting species, rather than by the distribution of $\Sigma$. It is also not clear what role the underlying dust disk plays in setting $r_\mathrm{in}$, as ultimately it is the sublimation of the solid debris disk particles that likely feeds the gaseous, line-emitting disk \citep{Rafikov1,Rafikov2,Garmilla,Metzger}. 

An eccentric, precessing disk scenario is not unique in its ability to produce periodic time-variabile emission line profiles, although its ability to reproduce variability on the observed timescales is promising. Testing the eccentric disk hypothesis would require self-consistent modeling of the emission from an eccentric disk \citep[e.g.,][]{Statler,Regaly}. It is important to emphasize that, in addition to determining the period of variability, our calculations also naturally provide the radial eccentricity profile of the disk (Figure \ref{fig:modes}). Any eccentric, rigidly precessing disk model used to reproduce the observed line profiles \citep{Cauley} should agree with these mode profiles. For example, the disk eccentricity should be largest near the inner edge and decrease with radius.\footnote{This is in part a consequence of the free inner boundary condition (\ref{eq:BC}) we have adopted, which allows the inner disk edge to be eccentric, in accordance with the observed emission line profiles.} Further, if $r_\mathrm{in}$ is small ($\lesssim 0.3 R_\odot$), the eccentricity should sharply decrease near the inner edge (cf.~\citealt{Cauley}), while for larger $r_\mathrm{in}$, the disk can be more uniformly eccentric. If $r_\mathrm{in}$ is sufficiently small, our calculations also impose an upper limit on the disk eccentricity in order to avoid orbit crossings (see Figure \ref{fig:emax}).

It is important to note that the qualitative behavior of $P_\mathrm{prec}$ is significantly affected by the inclusion of the 3D contribution in the eccentricity evolution equation (\ref{eq:E_evolution}). This term arises due to the variations of the vertical gravitational force along an eccentric orbit \citep{Ogilvie2008}, an effect omitted in \citet{go06}. If this term is neglected, pressure-dominated modes become retrograde ($\omega_\mathrm{prec} < 0$), while its inclusion results in prograde modes ($\omega_\mathrm{prec} > 0$). If the pressure-dominated modes were retrograde, then $\omega_\mathrm{prec}$ would cross zero and change sign near $r_\mathrm{crit}$ (since the GR-dominated modes are prograde), leading to a very small $|\omega_\mathrm{prec}|$, and therefore a very large $P_\mathrm{prec}$ ($\gtrsim 100$ yr). However, when the 3D effect is properly included, $\omega_\mathrm{prec}$ is always positive, and the very long periods associated with a zero-crossing of $\omega_\mathrm{prec}$ are excluded. We therefore find that the longest possible precession periods are $\approx 20$ years, implying that all gaseous debris disks should display signs of variability over years to decades (provided they are eccentric).

In this work we do not address the origin of the disk eccentricity, i.e., the physical mechanism behind its excitation, damping and saturation. Since WD debris disks are believed to be formed by the disruption of planetoids on nearly parabolic orbits, the disk eccentricity could be an artifact of the disruption. However, this ``primordial'' eccentricity would be viscously damped in several $10 - 100$s of years, depending on the value of the viscous $\alpha$ parameter, see equation (\ref{eq:t_nu}). Given that more than $50\%$ of known gaseous debris disks show variability likely related to their non-zero eccentricity, it is unlikely that the disk eccentricity is a transient phenomenon. 

This suggests that some eccentricity excitation mechanism must continually operate, so that the eccentricity is maintained by a balance between excitation and damping. One natural source of the eccentricity driving could be the interaction of the inner disk edge with the (inclined dipole) magnetic field of the WD that is responsible for the disk truncation.\footnote{We defer exploration of this possibility to future work.} Other possible mechanisms could include viscous overstability \citep{Kato,Lyubarskij}, or nontrivial aerodynamic coupling with the underlying particulate debris disk \citep{Rafikov2,Metzger}.

\acknowledgements

We thank Gordon Ogilvie, Boris G{\"a}nsicke, Jay Farihi, and Amy Bonsor for helpful comments and suggestions. Financial support for this study has been provided by NSF via grant AST-1409524 and NASA via grant 15-XRP15-2-0139.

\bibliographystyle{apj}
\bibliography{references}

\end{document}